\documentclass[useAMS,usenatbib,usedcolumn]{mn2e}
\def\HI{\ifmmode{\rm HI}\else{H\/{\sc i}}\fi}

\def\lsun{\ifmmode{{\mathrm L}_{\odot}}\else{L$_{\odot}$}\fi} 

\def\deg{\hbox{$^\circ$}}
\def\arcmin{\hbox{$^\prime$}}
\def\arcsec{\hbox{$^{\prime\prime}$}}

\def\magasas{\nobreak\mbox{$\;$mag$\,$arcsec$^{-2}$}}
\def\msun{\ifmmode{{\mathrm M}_{\odot}}\else{M$_{\odot}$}\fi} 
\def\msunpc2{\ifmmode{{\mathrm M}_{\odot} \, {\mathrm{pc}}^{-2}}\else{M$_{\odot} \, {\mathrm {pc}}^{-2}$}\fi}

\def\kms{\ifmmode{{\mathrm{km \, s^{-1}}}}\else{${\mathrm{km \, s^{-1}}}$}\fi} 

\newcommand{\OIII}{[O\/{\sc{iii}}]}

\usepackage[figuresright]{rotating}
\usepackage{lscape}
\usepackage{graphics}
\usepackage{epsfig}
\usepackage{multirow}
\usepackage{bigdelim}
\usepackage{bigstrut}

\title[Planetary nebula kinematics in NGC~1023]{Testing
  the nature of S0 galaxies using planetary nebula kinematics in
  NGC~1023}
\author[E.~Noordermeer et al.]
  {E.~Noordermeer,$^1$\thanks{email:edo.noordermeer@nottingham.ac.uk} 
   M.~R.~Merrifield,$^1$ L.~Coccato,$^2$ M.~Arnaboldi,$^3$ M.~Capaccioli,$^4$
  \newauthor  
   N.~G.~Douglas,$^5$ K.~C.~Freeman,$^6$ O.~Gerhard,$^2$ K.~Kuijken,$^7$
   F.~De~Lorenzi,$^2$ 
  \newauthor  
   N.~R.~Napolitano$^8$ and A.~J.~Romanowsky$^{9,10}$ \\ 
   $^1$University of Nottingham, School of Physics and Astronomy, University
       Park, NG7 2RD Nottingham, UK \\
   $^2$Max-Planck-Institut f\"ur Extraterrestrische Physik,
       Giessenbachstrasse,  85741 Garching, Germany \\
   $^3$European Southern Observatory, Karl-Schwarzschild-Strasse 2, 85748
       Garching, Germany \\ 
   $^4$Dipartimento di Fisica, Universit\`a ``Federico II'', Naples, Italy \\
   $^5$Kapteyn Astronomical Institute, University of Groningen, PO Box 800,
       9700 AV Groningen, The Netherlands \\
   $^6$Research School of Astronomy and Astrophysics, Australian National
       University, Canberra, Australia \\
   $^7$Leiden Observatory, Leiden University, PO Box 9513, 2300 RA Leiden, The
       Netherlands \\
   $^8$Istituto Nazionale di Astrofisica, Osservatorio Astronomico di
       Capodimonte, Via Moiariello 16, 80131 Naples, Italy \\
   $^9$UCO/Lick Observatory, University of California, Santa Cruz, CA 95064, 
       USA \\
   $^{10}$Departamento de F\'isica, Universidad de Concepci\'on, Casilla
       160-C, Concepci\'on, Chile}

\begin{document}

\date{accepted for publication in MNRAS; 5-12-2007}

\maketitle

\begin{abstract}
  We investigate the manner in which lenticular galaxies are formed by
  studying their stellar kinematics: an S0 formed from a fading spiral galaxy
  should display similar cold outer disc kinematics to its progenitor, while
  an S0 formed in a minor merger should be more dominated by random motions. 
  In a pilot study to attempt to distinguish between these scenarios,
  we have measured the planetary nebula (PN) kinematics of the nearby
  S0 system NGC~1023.  
  Using the Planetary Nebula Spectrograph, we have detected and measured the
  line-of-sight velocities of 204 candidate PNe in the field of this galaxy.
  Out to intermediate radii, the system displays the kinematics of a normal
  rotationally-supported disc system.  
  After correction of its rotational velocities for asymmetric drift, the
  galaxy lies just below the spiral galaxy Tully--Fisher relation, as one
  would expect for a fading system.  
  However, at larger radii the kinematics undergo a gradual but major
  transition to random motion with little rotation.  
  This transition does not seem to reflect a change in the viewing
  geometry or the presence of a distinct halo component, since the
  number counts of PNe follow the same simple exponential decline as
  the stellar continuum with the same projected disc ellipticity out
  to large radii.
  The galaxy's small companion, NGC~1023A, does not seem to be large enough to
  have caused the observed modification either.  
  This combination of properties would seem to indicate a complex evolutionary
  history in either the transition to form an S0 or in the past life of the
  spiral galaxy from which the S0 formed.  
  More data sets of this type from both spirals and S0s are needed in
  order to definitively determine the relationship between these types of 
  system.  
\end{abstract}

\begin{keywords}
galaxies: individual: NGC~1023 -- galaxies: individual: NGC~1023A -- galaxies:
elliptical and lenticular, cD -- galaxies: structure -- galaxies: kinematics
and dynamics -- galaxies: evolution 
\end{keywords}

%%%%%%%%%%%%%%%%%%%%%%%%%%%%%%%%%%%%%%%%%%%%%%%%%%%%%%%%%%%%%%%%%%%%%%%%%%%%%%%
%                                                                             %
%  1. Introduction                                                            %
%  \label{sec:introduction}                                                   %
%                                                                             %
%%%%%%%%%%%%%%%%%%%%%%%%%%%%%%%%%%%%%%%%%%%%%%%%%%%%%%%%%%%%%%%%%%%%%%%%%%%%%%%
\section{Introduction}
\label{sec:introduction}
Lenticular, or S0, galaxies make up about 25\% of all large galaxies
in the local Universe \citep{Dressler80}.  Their location in
\citeauthor{Hubble26}'s \citeyearpar{Hubble26} tuning fork diagram at
the apex between ellipticals and spirals emphasizes their importance
in understanding galaxy evolution, yet even the most basic question of
whether they are more closely related to the spiral systems on their
right or the ellipticals on their left remains unanswered. 
In the nearby Universe, S0s are more prevalent in galaxy clusters than in the
field, at the expense of a smaller fraction of spirals \citep{Dressler80}.
At higher redshifts, however, the S0 fraction in clusters is smaller, and the
spiral fraction correspondingly larger \citep{Dressler97, Couch98}, suggesting
that the latter are the progenitors of the lenticular galaxies observed in the
present-day Universe.  
How this transformation comes about, however, is still unclear. 
On the one hand, it is possible that S0s form when spiral galaxies lose their
gas through interactions with the intergalactic medium and thus simply stop
forming stars \citep{Quilis00}.  On the other, S0s could form from mergers:
while equal-mass mergers produce elliptical galaxies, the equivalent
unequal-mass mergers result in hybrid systems with both spheroidal and
disc-like components, which have lost most of their gas through starbursts
and tidal tails \citep{Bekki98, Bournaud04, Bournaud05}.

Distinguishing between these two scenarios is not straightforward. 
In principle, deep photometry of lenticulars should hold clues: in the
merger scenario, the central bulge extends outward into a diffuse
stellar halo, such that the light in the outer regions originates
mainly from a spheroidal component and the disc only dominates at
intermediate radii.  Indeed, a simple outward extrapolation of the
bulge intensity profiles of S0s seems to imply that this component
often wins out at large radii, with the disc only ever being important
at intermediate radii \citep{Seifert96}.  However, this behaviour
depends on the fitting functions and decomposition methods adopted,
how the sky background is flatfielded and subtracted, etc.; different
choices for these factors can give equally good fits, but with the
disc dominant out to the largest radii \citep{Laurikainen05}, as the
gas stripping scenario predicts.  It thus seems unlikely that
photometry alone will answer the question.

It is therefore perhaps more promising to look at the dynamics of
these systems, which may offer a clearer record of their life
histories.  If lenticulars are simply gas-stripped or quenched spirals, then
one would expect their stellar dynamics to be only modestly affected by the
transition. 
The gas stripping process may cause a final flickering of star formation in
the disc of the galaxy \citep{Quilis00}, and a modest degree of secular
evolution is needed to explain the larger bulge-to-disc ratio in lenticulars
compared to spiral galaxies (\citet{Christlein04}, although this result seems
to be contradicted by the results from \citet{Laurikainen05}).  
However, such effects are expected to play a r\^ole mainly in the central
parts of the galaxies. 
At large radii, the galaxies should largely keep the disc-like kinematics of a
spiral, with rotation dominating over random motions.  
On the other hand, if S0s are formed through mergers, simulations show that
the resulting systems are much `hotter', with velocity dispersions in the
outer regions as large as the rotational velocities
\citep{Bournaud04,Bournaud05}.  
In addition, the dynamics provide a measure of the mass 
distribution in these systems, allowing us to explore the scaling
relations between mass and luminosity.  Thus, for example, we can see
if lenticulars follow a similar Tully--Fisher relation to normal
spiral galaxies, as one might expect if the S0s are their descendants.

Unfortunately, results from previous dynamical studies have proved
somewhat inconclusive.  For example, \citet{Mathieu02} found that S0
galaxies obey a tight Tully--Fisher relation, but offset to fainter
magnitudes from the relation for spiral galaxies, suggesting that S0s
are formed from spirals that faded when they ceased forming new stars.
However, this result conflicts with the analysis of
\citet{Neistein99}, who found a huge scatter in the luminosity versus
rotation speed relation for S0s and, instead, found that the central
stellar velocity dispersion is a better predictor for the luminosity
via the fundamental plane relations, suggesting that S0s are more
closely related to ellipticals.  More recently, \citet{Bedregal06}
obtained much deeper absorption-line spectroscopy, which made the
determination of the rotation curve more reliable, and found a
relation between the offset from the spiral galaxy Tully--Fisher
relation and the age of S0s' stellar populations in the manner
expected if S0s are passively fading away from their former lives as
spirals.

However, even deep absorption-line spectroscopy is limited to the
brighter inner parts of S0 galaxies.  This limitation is unfortunate,
as much of the most useful dynamical information resides in the outer
parts of the system.  The Tully--Fisher analysis, for example,  relies
on getting out to large enough radii to determine the asymptotic form
of the rotation curve, and failure to do so can result in significant
biases in the results \citep{Noordermeer07c}.  In addition, the longer
dynamical timescales at larger radii means that any evidence as to a
system's life history will likely be better preserved at these radii.  

Fortunately, an alternative probe of stellar kinematics exists that
can operate at even the faintest surface brightness levels.  Planetary
nebulae (PNe) are simply low mass stars at the ends of their lives, so
their kinematics should faithfully trace the bulk stellar dynamics.
In addition, their strong emission features, particularly the 5007\AA\
\OIII\ line, means that they can be reliably identified and their
kinematics readily measured.  To exploit this resource, a specialized
instrument, the Planetary Nebula Spectrograph (PN.S), has been constructed. 
Mounted at the 4.2m William Herschel Telescope\footnote{The William
Herschel and Isaac Newton Telescopes are operated on the island of La Palma by 
the Isaac Newton Group in the Spanish Observatorio del Roque de los Muchachos
of the Instituto de Astrofisica de Canarias.}, the PN.S is designed to
identify PNe and measure their locations and velocities in a single
observation; details of the instrument and its operation can be found
in \citet{Douglas02}.

As a first test of the use of this instrument to understand the nature
of lenticular galaxies, we obtained an observation of the typical S0
system NGC~1023.  This object is the brightest of a group of 13
galaxies \citep{Tully80}, and, as one of the nearest large lenticular
galaxies at a distance of 11.4~Mpc \citep{Tonry01}, it offers an
obvious first target.  It also has a companion, NGC~1023A,
near the eastern edge of its disc at a projected distance of
9~kpc \citep{Barbon75}.
Based on the peculiar structure of neutral gas in the area it has been
suggested that these systems are closely interacting or even merging
\citep{Sancisi84, Capaccioli86}. 
However, after a careful analysis of an optical image of the NGC~1023/1023A
system, we show below that the companion is most likely too small to cause
significant disruptions in the dynamical state of the main galaxy.
Thus, the presence of the companion is probably of little influence for the
interpretation of the kinematical data we obtained. 

Absorption line spectroscopy for NGC~1023 has been presented by
various groups \citep{Simien97, Neistein99, Debattista02, Emsellem04}
and clearly shows that the galaxy is rotating regularly.  However,
these data underline the limitations of such conventional techniques,
since they reach to a maximum radius of only 100\arcsec, barely
outside the area dominated by the central bulge.  They thus do not
probe the regions likely to discriminate between the various scenarios
for S0 formation.  As we shall see below, the PN.S data reach to
approximately four times this radius, well into the disc-dominated
region and beyond.  

The remainder of this paper is structured as follows.  In
Section~\ref{sec:obs+data} we present the observations and describe
the basic data reduction.  In Section~\ref{sec:PNsystem}, we discuss
the spatial and velocity distribution of the detected PNe, and present
the method adopted to disentangle the objects belonging to NGC~1023
from those in NGC~1023A.  Section~\ref{sec:kinematics} explores the
dynamics of the main galaxy in more detail, derives the radial
profiles of rotation velocity and velocity dispersion, and places
NGC~1023 on the Tully--Fisher relation.  Finally, in
Section~\ref{sec:discussion}, we discuss the implications of our
results.

%%%%%%%%%%%%%%%%%%%%%%%%%%%%%%%%%%%%%%%%%%%%%%%%%%%%%%%%%%%%%%%%%%%%%%%%%%%%%%%
%                                                                             %
%  2. Observations and data reduction                                         %
%  \label{sec:obs+data}                                                       %
%                                                                             %
%%%%%%%%%%%%%%%%%%%%%%%%%%%%%%%%%%%%%%%%%%%%%%%%%%%%%%%%%%%%%%%%%%%%%%%%%%%%%%%
\section{Observations and data reduction}
\label{sec:obs+data}
\subsection{Planetary nebula data}
NGC~1023 was observed with the Planetary Nebula Spectrograph for a
total of 6.5 hours on the nights of 2006 October 23 and 27.  The
galaxy is too large to fit within the 10\arcmin\ PN.S field of view, so
the observation was split between two pointings (``West'' and
``East'') along the major axis of the galaxy, overlapped by about 4\arcmin.
Table~\ref{table:obs} provides a summary of the 
integration time and seeing. All exposures were made through the `AB'
filter, which has a central wavelength of 5026\AA\ and a FWHM bandpass
of 44\AA.  This wavelength coverage provides a velocity window for the
detection of PNe of -168 -- 2467~\kms, wide enough to detect all PNe
which are physically bound to NGC~1023, which has a systemic velocity
of $\sim 600 \, \kms$ (see below).
\begin{table}
 \centering
  \caption[PN.S observations of NGC~1023]
  {PN.S observations of NGC~1023.
    \label{table:obs}}

   \begin{tabular}{ccccc}
    \hline
    
    date & \multicolumn{2}{c}{exposure time} & seeing & photometric? \\
         & West & East &                              &              \\
         &  s   &   s  &                      \arcsec &              \\
    \hline

    23/10/2006 & 7200 & 7200 &                   1.0  & yes          \\
    27/10/2006 & 5400 & 3600 &                   1.4  & no           \\
    \hline
   \end{tabular}
\end{table}  

The data reduction followed the standard PN.S pipeline described in
detail in \citet{Douglas07}.  The only further enhancement to the
reduction involved the development of a procedure to help measure the
positions of PNe that were confused with the short continuum spectra
of stars in the spectrograph's field: by reflecting these ``star
trails'' about their centres in the undispersed direction, it proved
possible to subtract out the contribution of the stellar spectrum,
leaving just the point-like emission line from the PN.  

The observations for the East and West pointings were reduced
separately, and PN locations and velocities were derived independently from
each pointing.  Since the two pointings overlapped 
in the central regions of the galaxy, a significant number of objects
were detected in both, and the differences in their positions,
velocities and magnitudes provide a useful internal check on the
robustness of the instrument and the pipeline reduction.  The
positions, velocities and instrumental magnitudes for the 116 objects
found in both pointings have mean differences of 0.17\arcsec,
1.8~\kms\ and 0.05~mag, with an RMS scatter of 0.22\arcsec, 16.3~\kms\
and 0.15~mag respectively.  These errors are within the design
tolerance of the instrument: given the large spatial and velocity
scales of the structure under study, they will have no significant
impact on this dynamical study of NGC~1023.

In addition to the 116 objects detected in both pointings, 88 objects
were found in the individual pointings in the western and eastern
outskirts of the galaxy. The resulting final catalogue of 204 objects
is presented in Table~\ref{table:catalogue}.

\begin{figure}
 \centerline{\psfig{figure=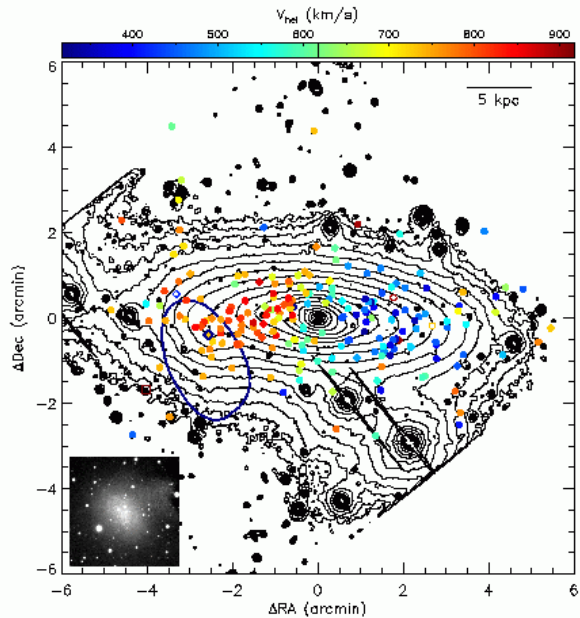,width=8.5cm}}
  \caption{The distribution of stellar light and planetary nebula in
    NGC~1023. Black contours show R-band isophotes, ranging from 14.5 to 
    25.5\magasas, in steps of 0.5\magasas. The solid bullets show the
    positions of the PNe, with the colours indicating the heliocentric 
    velocities. Open circles indicate the $>3 \sigma$-outliers from the
    tilted-ring analysis; the square symbol indicates the object which
    deviates more than 1000~\kms\ from the systemic velocity of the
    galaxy. The blue contour indicates the region where PNe are more likely to 
    belong to the dwarf companion galaxy NGC~1023A than to the main
    galaxy; the centre of the companion is indicated with the blue
    diamond. The 20 PNe in this region were removed from the kinematic 
    analysis. The inset shows a grayscale image of NGC~1023A, at the same
    scale as the main figure, created from the original image by subtracting a
    model of NGC~1023 (see text for details). The excess light on its western
    side is a result of imperfect subtraction of the main
    galaxy. \label{fig:overlay}} 
\end{figure}

\subsection{Imaging data}
We have supplemented the PN data with R- and B-band images obtained with the
Isaac Newton Telescope on 1995 December 25, using the Prime Focus Camera.  
The 300 second exposures were flatfielded and flux calibrated in the
conventional manner.  
A contour plot of the resulting R-band image is shown in
Figure~\ref{fig:overlay}, together with a colour coded representation of the
PN velocity field.

%%%%%%%%%%%%%%%%%%%%%%%%%%%%%%%%%%%%%%%%%%%%%%%%%%%%%%%%%%%%%%%%%%%%%%%%%%%%%%%
%                                                                             %
%  3. The Planetary Nebula system of NGC 1023                                 %
%  \label{sec:PNsystem}                                                       %
%                                                                             %
%%%%%%%%%%%%%%%%%%%%%%%%%%%%%%%%%%%%%%%%%%%%%%%%%%%%%%%%%%%%%%%%%%%%%%%%%%%%%%%
\section{The Planetary Nebula system of NGC 1023}
\label{sec:PNsystem}
Figure~\ref{fig:overlay} shows that the PN distribution generally follows
the optical image, except for the inner regions where the bright
continuum background inhibits the detection of PNe
\citep[cf.][]{Douglas07}.  Note that a significant number of objects
are detected out to the very faintest measured optical isophotes.  One
such object (ID~200, marked with the open square in Figure~\ref{fig:overlay})
has a highly discrepant velocity ($V_{\mathrm {hel}} = 1703 \, \kms$), and
is probably an unrelated background source.  This object has
accordingly been excluded from the dynamical analysis.  

\begin{figure}
 \centerline{\psfig{figure=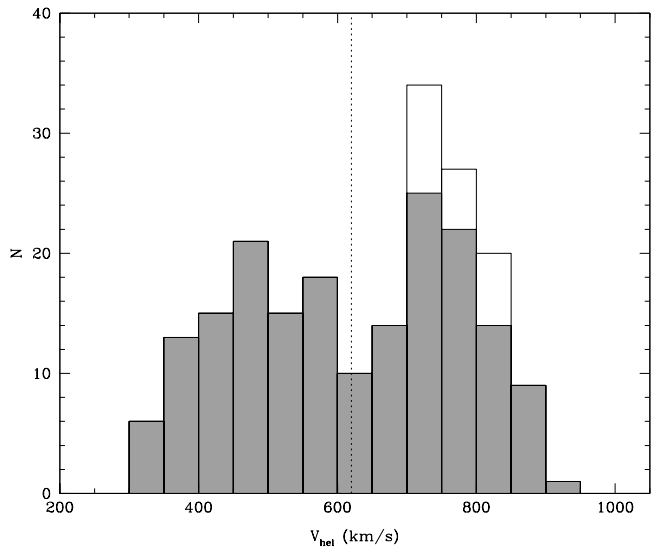,width=8.5cm}}
  \caption{The velocity distribution of the PNe system of NGC~1023. The grey
    histogram indicates the PNe in the main galaxy, whereas the white regions
    show the PNe which were assigned to the companion. The dotted line
    indicates the systemic velocity.\label{fig:veldist}} 
\end{figure}

The velocity distribution of the remaining objects is shown in
Figure~\ref{fig:veldist}.  The distribution shows a clear asymmetry,
with an excess of objects around $V_{\mathrm{hel}} = 750 \, \kms$.
This value coincides with the measured velocity for the dwarf
companion NGC~1023A \citep[$742 \pm 30 \kms$;][]{Capaccioli86} and, indeed,
inspection of Figure~\ref{fig:overlay} reveals a concentration of objects near
this velocity in the region where the isophotes show the distorted
signature of contamination by this object.  Inconveniently, the
rotation of NGC~1023 is such that one cannot cleanly separate the two
galaxies on the basis of their kinematics, since the mean velocities
of the systems are very similar in this region.  We therefore adopt
the procedure developed by \citet{Douglas07} to assign membership based on
local relative surface brightness contributions.  

\begin{figure}
 \centerline{\psfig{figure=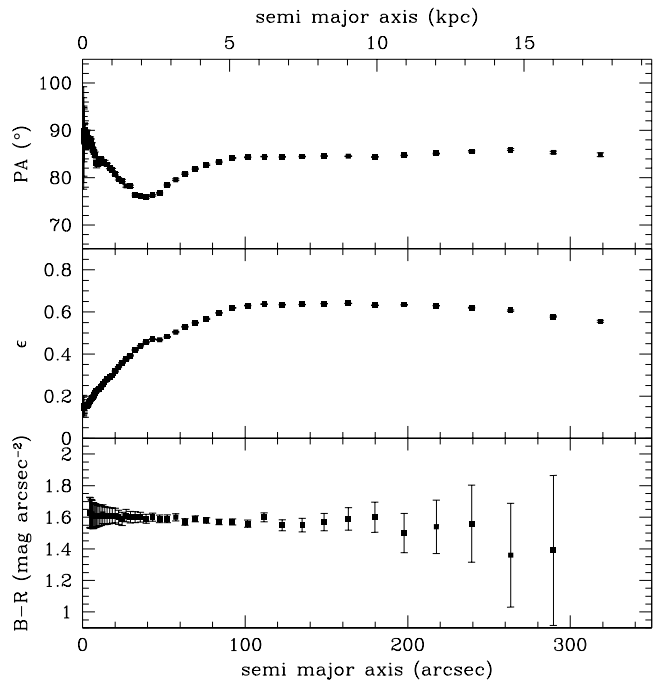,width=8.5cm}}
  \caption{Results from the isophotal analysis of NGC~1023. The top and
    middle panels show the fitted position angle (north through east) and
    ellipticity of the isophotes respectively. The bottom panel shows the B-R
    colour. \label{fig:isophots}}   
\end{figure}

\begin{figure}
 \centerline{\psfig{figure=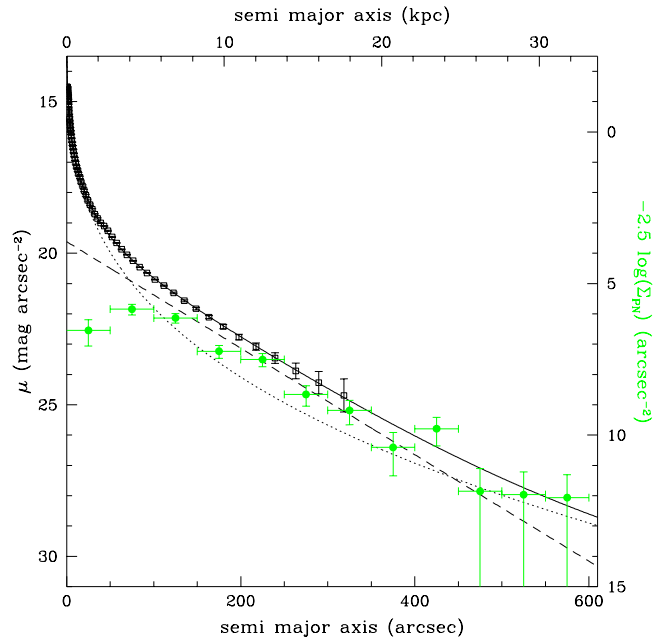,width=8.5cm}}
  \caption{Comparison of the radial PN density distribution in NGC~1023 (green
    filled circles) with the R-band photometric profile (black open squares).
    Both were measured on concentric ellipses with an ellipticity of 0.63 and
    a position angle of $85\deg$. Dotted and dashed lines indicate S\'ersic
    bulge and exponential disc profiles fitted to the R-band profile and the
    solid line shows the sum of the two components. The fitted bulge and disc
    have an effective radius and exponential scale length of 32 and 62\arcsec\
    (1.8 and 3.4~kpc) respectively, which are typical values for a lenticular
    galaxy of this luminosity \citep{Barway07}. \label{fig:surfden} }   
\end{figure}

We first study the optical morphology of NGC~1023 by masking out all
foreground stars, as well as the region contaminated by NGC~1023A, and fitting 
elliptical isophotes to the remaining parts of the image. The results from
this procedure are shown in the top and middle panel of
Figure~\ref{fig:isophots} and show that, after the exclusion of the companion,
NGC~1023 has a highly regular structure. 
In particular, the position angle and ellipticity of the 
galaxy are remarkably constant outside the bulge-dominated area and do not
show any signs of deviation from a simple, flat disc out to the outermost
radii. Note that the colour of the fitted isophotes is also constant with
radius within the errors (bottom panel in Figure~\ref{fig:isophots}). 
Based on the fitted isophotes and the derived intensity profile (see
Figure~\ref{fig:surfden}), we created a model image for NGC~1023,
interpolating over the masked regions. This model image was then subtracted
from the original, leaving the emission from NGC~1023A in the residual
image. A grayscale representation of its intensity distribution is shown in
the inset in Figure~\ref{fig:overlay}.   
Note that the companion is very small compared to NGC 1023 itself, with an
R-band main-to-companion luminosity ratio of about 40:1. 

We then followed the same method to create a model image for NGC~1023A, and
divided the two model images to calculate the ratio of flux from the two
galaxies at each location in the field, and hence to which system a PN at that
position is most likely to belong.  The only additional complication is that
we need to take into account the fact that NGC~1023A has a bluer colour than
NGC~1023, which increases the specific frequency of PNe.  Having allowed for
this effect using the relations between colour and specific frequency from
\citet{Hui93}, we can construct the contour at which a PN is equally likely to
be a member of either system, as shown in Figure~\ref{fig:overlay}. 

As the simplest approach to assigning membership, we associate the 20
PNe in this region with NGC~1023A, leaving 183 PNe in the final
NGC~1023 catalogue. 
As Figure~\ref{fig:veldist} shows, the 20 PNe associated with NGC~1023A have a
narrow velocity distribution with a full width of 130~\kms, centered around a
median velocity of 767~\kms. 
The over-all velocity distribution of the 183 remaining objects is symmetric
and has the two-horned structure characteristic of a rotating system.  
Clearly, the simple spatial cut applied here will result in some modest level
of misidentification, but experiments that excluded or included more objects
showed that the following results are not sensitive to this residual
contamination.
Furthermore, measurements of the average streaming velocities and
velocity dispersions on the approaching and receding halves of the galaxy
separately show that, after the exclusion of the NGC~1023A PNe, both sides are
consistent within the errorbars. 

In Figure~\ref{fig:surfden}, we compare the NGC~1023 PN number
densities with the R-band photometric profile.  Outside a radius of
about 100\arcsec, the PN number counts follow a similar exponential
distribution to the stellar light.  This good match lends support to
the assumption that the PNe are an unbiased tracer of the dominant
stellar components in such an old disc population \citep[see
also][]{Douglas07}.  At smaller radii, the PN counts become incomplete 
due to the difficulty of detecting them against the brighter parts of
the galaxy.  Given the nature of source detection in a slitless
spectrograph like the PN.S, there is also potential for kinematic as well
as photometric selection biases at the smallest radii where the
luminosity gradients are strongest \citep{Douglas07}, so we do
not use the PN kinematics in this region, instead basing our
dynamical analysis on the complementary absorption-line data that are
available at these higher surface brightnesses.  However, at larger
radii the completeness of the PN data means that we can reliably use
our measurements to study stellar kinematics all the way out to an
equivalent R-band surface brightness of about 28\magasas, or 8 disc
scale lengths, a region completely inaccessible to the conventional
absorption-line analysis.

%%%%%%%%%%%%%%%%%%%%%%%%%%%%%%%%%%%%%%%%%%%%%%%%%%%%%%%%%%%%%%%%%%%%%%%%%%%%%%%
%                                                                             %
%  4.The  kinematics of NGC~1023's PN system                                  %
%  \label{sec:kinematics}                                                     %
%                                                                             %
%%%%%%%%%%%%%%%%%%%%%%%%%%%%%%%%%%%%%%%%%%%%%%%%%%%%%%%%%%%%%%%%%%%%%%%%%%%%%%%
\section{Kinematic analysis}
\label{sec:kinematics}

\subsection{Rotational velocities and velocity dispersions}
\label{subsec:rotcur and veldisp}
To study the kinematic structure of NGC~1023, we first use a discrete
tilted ring analysis, which is in essence the equivalent for discrete
sources of the familiar tilted-ring method used for 2D gas velocity
fields \citep{Begeman87, Begeman89}.  
The sample of 183 PNe in NGC~1023 was divided into 11 radial bins, and the
velocities in each bin were fitted with a function of the form
\begin{equation}
  V_{\mathrm {obs}} = V_{\mathrm {sys}} + V_{\mathrm {rot}}
  \sin(i) \cos(\theta), 
\end{equation}
where $\theta$ is the azimuthal angle in the plane of the galaxy,
starting from the receding side on the major axis.  We assumed that
the disc of NGC~1023 has an inclination with respect to the line of
sight of $i = 71.5\deg$ (based on the ellipticity of the optical
isophotes and an assumed intrinsic thickness of the disc of 0.2) and a
position angle of $P\!A = 85\deg$ from north. 
We have tested the sensitivity of our results to the assumed values for the
orientation angles by doing additional fits with inclinations ranging from 65
to 75\deg, but found that the effects on the fitted velocities and velocity
dispersions (see below) were modest and did not affect our results
significantly. 
The uncertainties in $P\!A$ have negligible effect on the kinematic
measurements. 

\begin{figure}
 \centerline{\psfig{figure=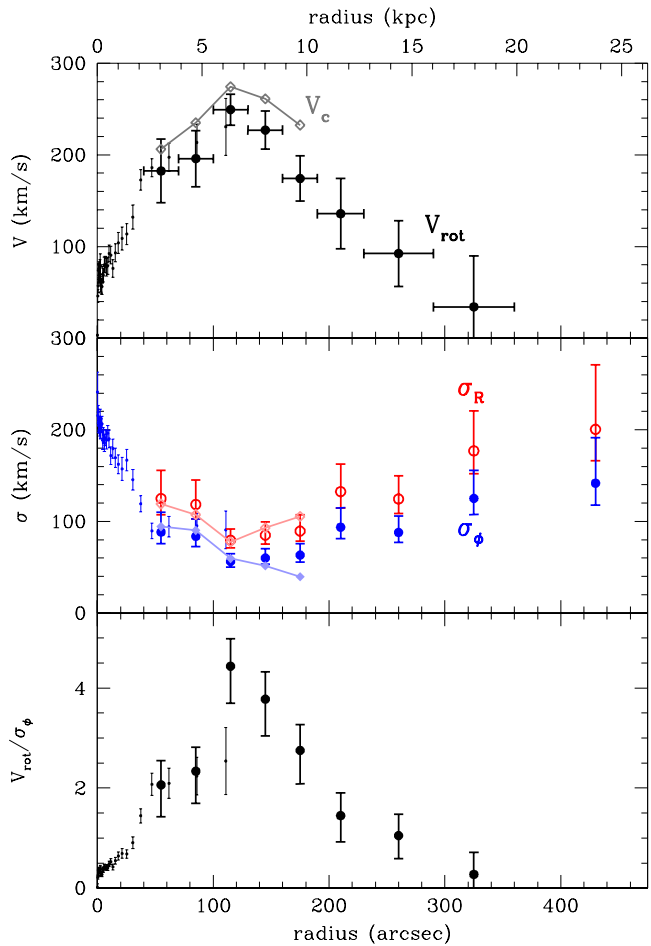,width=8.5cm}}
  \caption{Results from the kinematical analysis. The upper panel shows
  the mean streaming analysis. Small data points with errorbars are from the
  major-axis absorption-line data of \citet{Debattista02}. Large filled data
  points are from the tilted ring analysis of the PNe velocities; the
  errorbars indicate the formal fitting uncertainties. The connected grey
  diamonds show the rotation curve after iterative correction for asymmetric
  drift (see text for details). The middle panel shows the analysis of random
  motions. Small data points with errors are once again major-axis
  absorption-line data from \citet{Debattista02}. The open red symbols and
  large filled blue symbols are the radial and tangential components of
  velocity dispersion respectively, as inferred from the tilted-ring fit to
  the PNe data; the errorbars show, again, the formal errors. The connected
  diamonds show the corresponding values after the iterative correction for
  asymmetric drift. The bottom panel shows the ratio between rotational motion
  and the azimuthal component of the velocity dispersion. \label{fig:RC}}
\end{figure}

In the first instance, we let the systemic velocity $V_{\mathrm {sys}}$ vary
freely from ring to ring.  
The values found from that fit were then averaged to find a mean of 617~\kms,
which was subsequently fixed as the global systemic velocity for the entire
galaxy. 
In order to account for the possibility of contamination by unrelated objects
such as background galaxies, a simple 3~sigma-clipping criterion was applied
to exclude outliers from the fit; the 3 PNe with discrepant velocities which
we identified in this manner are indicated with the open symbols in
Figure~\ref{fig:overlay}.  
The final resulting mean streaming velocity as a function of ring radius is
shown in the upper panel of Figure~\ref{fig:RC}.

In addition to the rotational velocities of the PNe in each ring, we
also measure the residual velocities around the fits.  These residuals
constrain the velocity dispersions along the principal axes of the
system, with the measured line-of-sight dispersion at a given location
in the disc given by:
\begin{equation}
  \sigma_{\mathrm {los}}^2 = \sigma_R^2 \: \sin^2 \! i \: \sin^2 \! \theta + 
                          \sigma_\phi^2 \: \sin^2 \! i \: \cos^2 \! \theta + 
                             \sigma_z^2 \: \cos^2 \! i,
  \label{eq:sigmaeff}
\end{equation}
where $\sigma_R$, $\sigma_\phi$ and $\sigma_z$ are the components of
dispersion along the cylindrical polar axes. NGC~1023 is sufficiently close to
edge-on and the $z$-component of random velocity sufficiently small that
$\sigma_z$ makes a negligible contribution to the observed dispersion.  In
principle, the remaining two components can be disentangled by  measuring
$\sigma_{\mathrm {los}}$ as a function of azimuth.  However, with only $\sim
20$ PNe per radial bin, there is not enough signal to derive these
components individually.  Instead, we solve for a single component
under the assumption that the other component is coupled to it via the
epicyclic approximation,   
\begin{equation}
  \frac{\sigma_\phi^2}{\sigma_R^2} = 
  \frac{1}{2} \left( 1 + \frac{d (\ln V_c)}{d (\ln R)} \right).
  \label{eq:epicycle}
\end{equation}
\citep{Binney87}.  Clearly, this approximation breaks down if the random
velocities become too large, but it offers a useful model for starting
to understand the dynamics of the disc-dominated parts of this system.

There is one further complication in that the circular speed, $V_c$,
for this system is not known since the observed rotation speed of the
stars is lowered due to asymmetric drift (see
Section~\ref{subsec:asymmetricdrift}).  However, as a first
approximation it is probably not too unreasonable to assume a flat
rotation curve outside the very central regions, so that the derivative in
equation~\ref{eq:epicycle} vanishes at the radii of the PNe and $\sigma_R /
\sigma_\phi = \sqrt{2}$. 
With this enforced coupling, we obtain the velocity dispersion profiles for
the individual components shown as data points in the middle panel of
Figure~\ref{fig:RC}.

Several points are immediately apparent from even this approximate
analysis.  First, it is clear that the kinematics inferred from the PNe match
up very well to those inferred from the absorption line data of
\citet{Debattista02}, also shown in Figure~\ref{fig:RC}.  This match further
confirms that the two types of data sample 
the same underlying kinematic tracer and underlines the
complementarity between the data sets, with the PNe taking over at
just the point where the absorption-line data give out.  Second, it is
immediately clear that the absorption-line data do not tell the full
story with this galaxy, as out to the largest radius that they trace
one obtains a picture of a well-behaved system with monotonically
declining dispersion profiles and rising rotation profile.  However,
at the much larger radii traced by the PNe these trends unexpectedly
reverse entirely, and the velocity dispersions increase rapidly at the
expense of a sharply dropping rotational velocity profile.  As a
result, the PN system becomes dominated again by random motions
outside a radius of about 250\arcsec (14~kpc), with rotation only
being dominant at intermediate radii (between 1.7 and 13~kpc).

Before continuing our analysis, we must ask the question whether geometric
effects can explain this peculiar behaviour.  
If the outer disc were significantly flared or warped towards edge-on, the
line of sight would sample a large range of radii, effectively lowering the
measured $V_{\mathrm {rot}}$ and increasing $\sigma_\phi$. 
Although the regularity of the outer isophotes of NGC~1023 (see
Figure~\ref{fig:isophots}) seems to argue against such effects, we have
nevertheless carried out simple simulations to study the effects of thickened
discs and varying viewing angles on the observed velocity distributions. 
We have found that, although line-of-sight integration effects can indeed
increase the velocity dispersions, the effect is always at least an order of
magnitude smaller than the velocity amplitude of the rotation curve. We have
not been able to reproduce the observed velocity dispersions other than by
assuming that they are intrinsic. 
Furthermore, although the line-of-sight integration effects did indeed lower
the mean streaming velocities along the major axis of the system, the mean
velocities were often increased at other position angles, such that the
azimuthally averaged rotation velocity remained largely unchanged.  
\begin{figure}
 \centerline{\psfig{figure=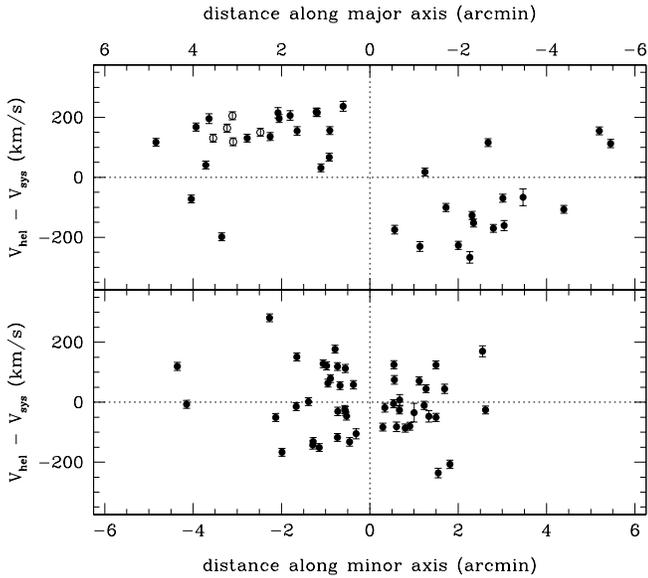,width=8.5cm}}
  \caption{PNe position-velocity diagrams along the major (top panel) and
    minor (bottom) axes. Each panel shows the velocities of all objects that
    lie within 20\deg\ in $\theta$ of the respective axes. Open symbols
    indicate PNe in the region contaminated by NGC~1023A (i.e. objects inside
    the blue contour in Figure~\ref{fig:overlay}). \label{fig:posvel}} 
\end{figure}

As a final check, we show in Figure~\ref{fig:posvel} the velocities of
individual PNe within 20\deg\ of the major and minor axes of NGC~1023. 
The increase in velocity dispersion with radius and the associated decrease
in average streaming motion along the major axis are clearly visible. 
Note in particular the counter-rotating PNe on both sides of the major
axis; such objects would not exist in a flared or warped, but otherwise
regularly rotating, disc.

\subsection{Asymmetric drift and the circular velocity curve}
\label{subsec:asymmetricdrift}  
To proceed further with our analysis, for example to study the
location of NGC~1023 on the Tully--Fisher relation, we must derive the
rotation curve.  To do so, we must correct the measured mean rotation
of the stellar component for asymmetric drift via the equation
\begin{equation}
  V_c^2 = V_{\mathrm {rot}}^2 + \sigma_\phi^2 - 
          \sigma_R^2 \left( 1 + \frac{d (\ln \nu)}{d (\ln R)} \right) -
          R \frac{d \sigma_R^2}{d R},
  \label{eq:asymdrift}
\end{equation}
where we have assumed that the tilt term, $d (\overline{V_RV_z}) /
dz$, is negligible \citep{Binney87}.  This equation is only valid when
the motions are dominated by rotation and cannot be applied to systems
dominated by random motions, so we only apply it to the points at intermediate
radii ($40\arcsec < R < 200\arcsec$) where the mean streaming motions exceed
the random velocities (see Figure~\ref{fig:RC}).  
At these radii, the density of the kinematic tracer, $\nu(r)$, is well
described by a simple exponential with scale-length $h = 62\arcsec$ (see
Figure~\ref{fig:surfden}), so we adopt that functional form for this
analysis. 

Now, we have three coupled equations, \ref{eq:sigmaeff} --
\ref{eq:asymdrift}, for the three unknowns, $V_c$, $\sigma_\phi$ and
$\sigma_R$, which we can solve in an iterative fashion.  Starting
with the data points from Figure~\ref{fig:RC}, we use
equation~\ref{eq:asymdrift} to obtain an estimate for $V_c$.  This
rotation curve is then used to re-evaluate the ratio between the
azimuthal and radial velocity dispersions via
equation~\ref{eq:epicycle} and new dispersion profiles are calculated,
which in turn are used to refine the estimate for the rotation curve.
This process converges rapidly, and results in the final dispersion
profiles and rotation curve shown connected by solid lines in
Figure~\ref{fig:RC}.

It is apparent that the refinement of the rotation curve does not have
much impact on the inferred shape of the velocity ellipsoid, beyond a
stretching of its shape in the radial direction at large radii.  The
correction for asymmetric drift does, however, result in a significant
difference between the mean streaming speed and the inferred rotation
curve.  While the mean streaming velocity was found to drop faster
than a Keplerian decline, the rotation curve drops more slowly, as
is physically required. Indeed, between 100 and 200\arcsec, the variation in
circular speed with radius is relatively modest and, given the large errorbars
on the original uncorrected data points, the rotation curve is consistent with
being flat. 
\begin{figure}
 \centerline{\psfig{figure=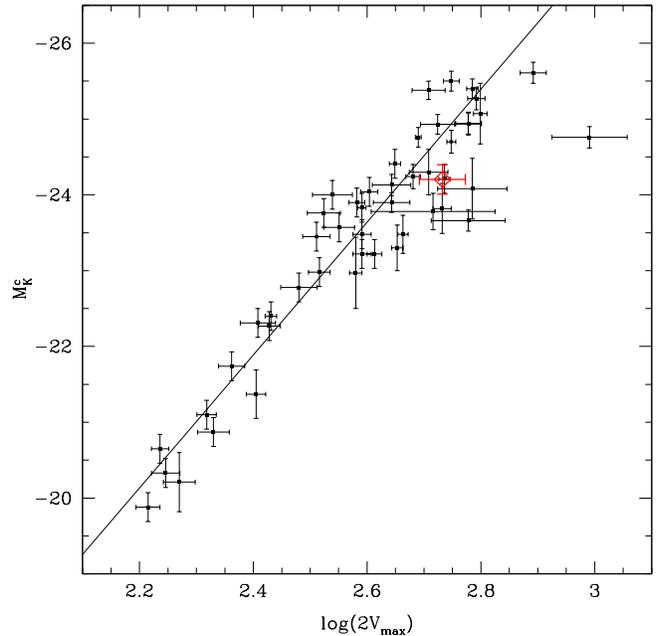,width=8.5cm}} 
  \caption{The location of NGC~1023 on the K-band Tully--Fisher
    relation. Small black data points show the Tully--Fisher relation from
    \citet{Noordermeer07c}, with the solid line showing the best linear
    fit. The location of NGC~1023 is indicated with the large red
    diamond. \label{fig:TF}}  
\end{figure}

\subsection{The location of NGC~1023 on the Tully-Fisher relation}
\label{subsec:TF}
From the rotation curve shown in Figure~\ref{fig:RC}, we estimate a maximum
circular speed of $V_{\mathrm {c,max}} = 270 \pm 25 \, \kms$, which we use to
place NGC~1023 on the K-band Tully--Fisher relation. For its luminosity, we
adopt the total apparent magnitude as measured by 2MASS\footnote{2MASS
(http://www.ipac.caltech.edu/2mass/) is a joint project of the
University of Massachusetts and the Infrared Processing and Analysis
Center/California Institute of Technology, funded by the National
Aeronautics and Space Administration and the National Science
Foundation.}, and follow the same conversions as in
\citet{Noordermeer07c} to derive the absolute magnitude. This figure
is marginally inflated since the 2MASS data have not subtracted out
the contribution from NGC~1023A.  However, from the R-band image in
Figure~\ref{fig:overlay}, we know that this contribution is at most
0.02~magnitudes, which is not large enough to cause any significant
shift.  Using these values, we place NGC~1023 on the Tully--Fisher
relation shown in Figure~\ref{fig:TF}.  

It is clear from this figure that NGC~1023's location is entirely consistent
with what one would expect for a normal disc galaxy: its offset below the mean
relation ($\approx -0.6$~mag) is only 1.5 times the scatter as measured by 
\citet{Noordermeer07c} and the uncertainties in $V_{\mathrm {c,max}}$ can
account for most of this offset. 
The offset can, alternatively, also be explained if NGC~1023 formed from a
normal star-forming spiral galaxy about 1~Gyr ago, and has been fading
passively since \citep{Bedregal06}.

%%%%%%%%%%%%%%%%%%%%%%%%%%%%%%%%%%%%%%%%%%%%%%%%%%%%%%%%%%%%%%%%%%%%%%%%%%%%%%%
%                                                                             %
%  5. Discussion and conclusions                                              %
%  \label{sec:discussion}                                                     %
%                                                                             %
%%%%%%%%%%%%%%%%%%%%%%%%%%%%%%%%%%%%%%%%%%%%%%%%%%%%%%%%%%%%%%%%%%%%%%%%%%%%%%%
\section{Discussion and conclusions}
\label{sec:discussion}
In this paper, we have set out to show that dynamical studies of the
planetary nebulae in S0 galaxies can shed some light on the origins of
these systems through a pilot study of the archetypal lenticular
system NGC~1023.  However, as so often in astronomy, the picture that
emerges from these data is somewhat complex.

At small radii, the absorption-line spectra show a well-behaved
dynamical structure typical of a system that is bulge-dominated (i.e.\
kinematically `hot') at its centre, with a transition to rotation-dominated
disc kinematics at larger radii. 
This picture is confirmed and extended by the PN data to a reasonably flat
maximum in the rotation curve and a continuing decline in the velocity
dispersions.
Combining these data with the K-band luminosity places NGC~1023 a little below
the mean Tully--Fisher relation for spiral galaxies, as one would expect for a
passively fading stellar population.  
The data in these regions fit very nicely with a picture in which this
galaxy evolved from a normal spiral system that shut down its star
formation after losing its gas.  

The largest radii probed by the PNe uncover a completely different story and
contradict the `fading spiral' hypothesis.
Outside four disc scale-lengths, the kinematics become completely dominated by
random motions. 
Having excluded the possibility that this behaviour is caused by a
change in viewing geometry (e.g.\ a strongly warped or flared disc),
this change suggests that we are seeing a transition from a disc
population to the kind of hot stellar halo that might be produced by a
minor merger.  
If such a merger stripped the gas from the galaxy, it would have caused the
end of star formation as well and the subsequent fading of the stellar
population would again be consistent with the observed offset from the mean
Tully-Fisher relation. 
There are, however, no signs of any accompanying change in the photometry at
these radii, as both the stellar light and PN number counts seem to be
well reproduced by a single exponential profile (see
Figure~\ref{fig:surfden}).  
There is little indication either of the continuum image or PN distribution
becoming rounder at the transition radius; instead, the outer isophotes appear
as elongated as the ones at intermediate radii (see
Figure~\ref{fig:isophots}).   
Moreover, the observed dynamical structure seems to be inconsistent with what
is predicted in the minor merger simulations.  
Mergers with the mass ratios required to transform a spiral galaxy into a
lenticular \citep[5:1 -- 10:1,][]{Bournaud05} appear to affect the
dynamical structure of the remnant at all radii, such that they are
not rotationally-dominated anywhere. 
In particular, they do not reproduce the region with $V_{\mathrm
  {rot}} / \sigma_R > 3$ observed at intermediate radii in NGC~1023. 
Furthermore, the simulations predict a flat, or slowly declining,
velocity dispersion profile in the outer regions, in contrast with the
observed sharp increase in the case of NGC~1023. 
Note that mergers between spirals and ellipticals do not reproduce
the peculiar kinematics of NGC~1023 either \citep[e.g.][]{DiMatteo07}. 

So, to sum up, the fading spiral scenario fits with the properties of
NGC~1023 at small radii but fails to explain its strange kinematic
structure at large radii, while the minor merger scenario fits better
with the data at large radii but fails to explain the presence of a
rotation-dominated disc at small radii.

One possibility is that the disturbed kinematics reflect an ongoing
interaction with NGC~1023A, but this seems rather unlikely. 
As the companion is somewhat bluer than the main galaxy, the stellar mass
ratio of main-to-companion galaxy is likely to be more extreme than the R-band
luminosity ratio, which at 40:1 is already large. 
Even though NGC~1023A may be dark matter dominated, it seems implausible on
energetic grounds that such a minor merger could have such a profound
influence on the main galaxy's kinematics. 
Simulations of minor mergers show that they can heat up galactic discs and
produce, in some cases, velocity dispersions up to 100~\kms\
\citep{Bournaud04}, but this requires mass ratios as small as 10:1 or even
5:1. 
Mergers with smaller satellites increase the velocity dispersions by at most a
few tens of \kms\ \citep{Quinn93, Velazquez99}, in stark contrast with the
dispersions in the outer parts of NGC~1023 ($\approx 200$~\kms). 
Moreover, the fact that the optical morphology of the companion seems
relatively undisturbed (see the inset in Figure~\ref{fig:overlay}) seems hard
to explain if the interaction were responsible for the large disturbance in
the main galaxy. 
Finally, even if this interaction were in some way significant, it would not
explain why star formation in the inner parts of the main galaxy, which have
not been affected dynamically, has shut down. 
So we conclude that NGC~1023A cannot be central to the question of how
NGC~1023 was transformed into an S0 galaxy originally. 

A more interesting scenario was presented by \citet{Burkert05}, who reported
the discovery of a ring of diffuse star clusters in the disk of NGC~1023,
which rotate but with large random motions, and suggested that NGC~1023 has
undergone a head-on collision with a companion galaxy. 
\citet{Elmegreen07} showed that the star formation triggered by such an event
will naturally lead to the formation of the kind of diffuse clusters observed
in this galaxy.  
The maximum in the $V/\sigma$-profile in Figure~\ref{fig:RC} coincides with
the ring of clusters, which suggests that the former was produced in the
collision event as well. 
Note, however, that we do not observe a ring-like structure in the
distribution of PNe (see Figure~\ref{fig:surfden}), which makes the connection
somewhat speculative. 

One other remaining plausible scenario is that the outer kinematics of
NGC~1023 pre-date its transformation to an S0.  If the progenitor spiral
already had peculiar kinematics when it was stripped of its gas, then they
would have remained imprinted in the final lenticular system. Similarly, if
the parent galaxy in a minor merger has a constant velocity dispersion profile
at large radii, rather than the declining profile usually assumed in
simulations \citep[e.g.][]{Bournaud05}, than the end-product may well have a
rising dispersion profile instead of a flat one. Very few spiral galaxies have
had their stellar kinematics probed out to beyond three scale-lengths where
the peculiarities begin, so we do not know how likely such a scenario might
be. We do, however, already have some initial indications that random motions
may be more prevalent in the outskirts of stellar discs: in their study of the
PN kinematics of M31, \citet{Merrett06} found that outside two disc
scale-lengths in this system the expected decline in stellar velocity
dispersion with radius ceased, while \citet{Ciardullo04} and
\citet{Herrmann05} found similar results for M33 and M83. 

Ultimately, the conclusion here must be that even apparently simple-looking
systems like S0s can be dynamically quite complex, presumably reflecting the
complexities of the reality of galaxy evolution. 
Based on the single system presented in this study, we cannot make definitive
statements about which processes are most important in the transformation of
spiral galaxies into lenticulars. 
To disentangle the complexities in this transformation, we must not only look
in detail at all the clues provided by each individual system, but we must
also study significant numbers of both types of galaxies.

\section*{Acknowledgements}
We would like to thank Andi Burkert for valuable comments. 
AJR was supported by the FONDAP Center for Astrophysics CONICYT 15010003 and
by the National Science Foundation Grant AST-0507729. MRM is currently funded
by an STFC Senior Fellowship. 
We thank the referee, Frederic Bournaud, for a thorough review and useful
suggestions.

\bibliographystyle{mn2e}
\bibliography{abbrev,refs}

\begin{thebibliography}{}

\bibitem[\protect\citeauthoryear{{Barbon} \& {Capaccioli}}{{Barbon} \&
  {Capaccioli}}{1975}]{Barbon75}
{Barbon} R.,  {Capaccioli} M.,  1975, \aap, 42, 103

\bibitem[\protect\citeauthoryear{{Barway}, {Kembhavi}, {Wadadekar}, {Ravikumar}
  \& {Mayya}}{{Barway} et~al.}{2007}]{Barway07}
{Barway} S.,  {Kembhavi} A.,  {Wadadekar} Y.,  {Ravikumar} C.~D.,    {Mayya}
  Y.~D.,  2007, \apjl, 661, L37

\bibitem[\protect\citeauthoryear{{Bedregal}, {Arag{\'o}n-Salamanca} \&
  {Merrifield}}{{Bedregal} et~al.}{2006}]{Bedregal06}
{Bedregal} A.~G.,  {Arag{\'o}n-Salamanca} A.,    {Merrifield} M.~R.,  2006,
  \mnras, 373, 1125

\bibitem[\protect\citeauthoryear{Begeman}{Begeman}{1987}]{Begeman87}
Begeman K.,  1987, PhD thesis, Rijksuniversiteit Groningen

\bibitem[\protect\citeauthoryear{{Begeman}}{{Begeman}}{1989}]{Begeman89}
{Begeman} K.~G.,  1989, \aap, 223, 47

\bibitem[\protect\citeauthoryear{{Bekki}}{{Bekki}}{1998}]{Bekki98}
{Bekki} K.,  1998, \apjl, 502, L133

\bibitem[\protect\citeauthoryear{{Binney} \& {Tremaine}}{{Binney} \&
  {Tremaine}}{1987}]{Binney87}
{Binney} J.,  {Tremaine} S.,  1987, {Galactic dynamics}.
Princeton, NJ, Princeton University Press, 1987

\bibitem[\protect\citeauthoryear{{Bournaud}, {Combes} \& {Jog}}{{Bournaud}
  et~al.}{2004}]{Bournaud04}
{Bournaud} F.,  {Combes} F.,    {Jog} C.~J.,  2004, \aap, 418, L27

\bibitem[\protect\citeauthoryear{{Bournaud}, {Jog} \& {Combes}}{{Bournaud}
  et~al.}{2005}]{Bournaud05}
{Bournaud} F.,  {Jog} C.~J.,    {Combes} F.,  2005, \aap, 437, 69

\bibitem[\protect\citeauthoryear{{Burkert}, {Brodie} \& {Larsen}}{{Burkert}
  et~al.}{2005}]{Burkert05}
{Burkert} A.,  {Brodie} J.,    {Larsen} S.,  2005, \apj, 628, 231

\bibitem[\protect\citeauthoryear{{Capaccioli}, {Lorenz} \&
  {Afanasjev}}{{Capaccioli} et~al.}{1986}]{Capaccioli86}
{Capaccioli} M.,  {Lorenz} H.,    {Afanasjev} V.~L.,  1986, \aap, 169, 54

\bibitem[\protect\citeauthoryear{{Christlein} \& {Zabludoff}}{{Christlein} \&
  {Zabludoff}}{2004}]{Christlein04}
{Christlein} D.,  {Zabludoff} A.~I.,  2004, \apj, 616, 192

\bibitem[\protect\citeauthoryear{{Ciardullo}, {Durrell}, {Laychak}, {Herrmann},
  {Moody}, {Jacoby} \& {Feldmeier}}{{Ciardullo} et~al.}{2004}]{Ciardullo04}
{Ciardullo} R.,  {Durrell} P.~R.,  {Laychak} M.~B.,  {Herrmann} K.~A.,  {Moody}
  K.,  {Jacoby} G.~H.,    {Feldmeier} J.~J.,  2004, \apj, 614, 167

\bibitem[\protect\citeauthoryear{{Couch}, {Barger}, {Smail}, {Ellis} \&
  {Sharples}}{{Couch} et~al.}{1998}]{Couch98}
{Couch} W.~J.,  {Barger} A.~J.,  {Smail} I.,  {Ellis} R.~S.,    {Sharples}
  R.~M.,  1998, \apj, 497, 188

\bibitem[\protect\citeauthoryear{{Debattista}, {Corsini} \&
  {Aguerri}}{{Debattista} et~al.}{2002}]{Debattista02}
{Debattista} V.~P.,  {Corsini} E.~M.,    {Aguerri} J.~A.~L.,  2002, \mnras,
  332, 65

\bibitem[\protect\citeauthoryear{{Di Matteo}, {Combes}, {Melchior} \&
  {Semelin}}{{Di Matteo} et~al.}{2007}]{DiMatteo07}
{Di Matteo} P.,  {Combes} F.,  {Melchior} A.~.,    {Semelin} B.,  2007, ArXiv
  e-prints: 0710.1293

\bibitem[\protect\citeauthoryear{{Douglas}, {Arnaboldi}, {Freeman}, {Kuijken},
  {Merrifield}, {Romanowsky}, {Taylor}, {Capaccioli}, {Axelrod}, {Gilmozzi},
  {Hart}, {Bloxham} \& {Jones}}{{Douglas} et~al.}{2002}]{Douglas02}
{Douglas} N.~G.,  {Arnaboldi} M.,  {Freeman} K.~C.,  {Kuijken} K.,
  {Merrifield} M.~R.,  {Romanowsky} A.~J.,  {Taylor} K.,  {Capaccioli} M.,
  {Axelrod} T.,  {Gilmozzi} R.,  {Hart} J.,  {Bloxham} G.,    {Jones} D.,
  2002, \pasp, 114, 1234

\bibitem[\protect\citeauthoryear{{Douglas}, {Napolitano}, {Romanowsky},
  {Coccato}, {Kuijken}, {Merrifield}, {Arnaboldi}, {Gerhard}, {Freeman},
  {Merrett}, {Noordermeer} \& {Capaccioli}}{{Douglas} et~al.}{2007}]{Douglas07}
{Douglas} N.~G.,  {Napolitano} N.~R.,  {Romanowsky} A.~J.,  {Coccato} L.,
  {Kuijken} K.,  {Merrifield} M.~R.,  {Arnaboldi} M.,  {Gerhard} O.,  {Freeman}
  K.~C.,  {Merrett} H.~R.,  {Noordermeer} E.,    {Capaccioli} M.,  2007, \apj,
  664, 257

\bibitem[\protect\citeauthoryear{{Dressler}}{{Dressler}}{1980}]{Dressler80}
{Dressler} A.,  1980, \apj, 236, 351

\bibitem[\protect\citeauthoryear{{Dressler}, {Oemler}, {Couch}, {Smail},
  {Ellis}, {Barger}, {Butcher}, {Poggianti} \& {Sharples}}{{Dressler}
  et~al.}{1997}]{Dressler97}
{Dressler} A.,  {Oemler} A.~J.,  {Couch} W.~J.,  {Smail} I.,  {Ellis} R.~S.,
  {Barger} A.,  {Butcher} H.,  {Poggianti} B.~M.,    {Sharples} R.~M.,  1997,
  \apj, 490, 577

\bibitem[\protect\citeauthoryear{{Elmegreen}}{{Elmegreen}}{2007}]{Elmegreen07}
{Elmegreen} B.~G.,  2007, ArXiv e-prints: 0710.5788

\bibitem[\protect\citeauthoryear{{Emsellem}, {Cappellari}, {Peletier},
  {McDermid}, {Bacon}, {Bureau}, {Copin}, {Davies}, {Krajnovi{\' c}},
  {Kuntschner}, {Miller} \& {Tim de Zeeuw}}{{Emsellem}
  et~al.}{2004}]{Emsellem04}
{Emsellem} E.,  {Cappellari} M.,  {Peletier} R.~F.,  {McDermid} R.~M.,  {Bacon}
  R.,  {Bureau} M.,  {Copin} Y.,  {Davies} R.~L.,  {Krajnovi{\' c}} D.,
  {Kuntschner} H.,  {Miller} B.~W.,    {Tim de Zeeuw} P.,  2004, \mnras, 352,
  721

\bibitem[\protect\citeauthoryear{{Herrmann} \& {Ciardullo}}{{Herrmann} \&
  {Ciardullo}}{2005}]{Herrmann05}
{Herrmann} K.~A.,  {Ciardullo} R.,  2005, in {Szczerba} R.,  {Stasinska} G.,
  {Gorny} S.~K.,  eds, Planetary Nebulae as Astronomical Tools Vol.~804 of
  American Institute of Physics Conference Series, {Planetary Nebula Studies of
  Face-On Spiral Galaxies: Is the Disk Mass-to-Light Ratio Constant?}.
pp 341--344

\bibitem[\protect\citeauthoryear{{Hubble}}{{Hubble}}{1926}]{Hubble26}
{Hubble} E.~P.,  1926, \apj, 64, 321

\bibitem[\protect\citeauthoryear{{Hui}, {Ford}, {Ciardullo} \& {Jacoby}}{{Hui}
  et~al.}{1993}]{Hui93}
{Hui} X.,  {Ford} H.~C.,  {Ciardullo} R.,    {Jacoby} G.~H.,  1993, \apj, 414,
  463

\bibitem[\protect\citeauthoryear{{Laurikainen}, {Salo} \& {Buta}}{{Laurikainen}
  et~al.}{2005}]{Laurikainen05}
{Laurikainen} E.,  {Salo} H.,    {Buta} R.,  2005, \mnras, 362, 1319

\bibitem[\protect\citeauthoryear{{Mathieu}, {Merrifield} \&
  {Kuijken}}{{Mathieu} et~al.}{2002}]{Mathieu02}
{Mathieu} A.,  {Merrifield} M.~R.,    {Kuijken} K.,  2002, \mnras, 330, 251

\bibitem[\protect\citeauthoryear{{Merrett}, {Merrifield}, {Douglas}, {Kuijken},
  {Romanowsky}, {Napolitano}, {Arnaboldi}, {Capaccioli}, {Freeman}, {Gerhard},
  {Coccato}, {Carter}, {Evans}, {Wilkinson}, {Halliday} \& {Bridges}}{{Merrett}
  et~al.}{2006}]{Merrett06}
{Merrett} H.~R.,  {Merrifield} M.~R.,  {Douglas} N.~G.,  {Kuijken} K.,
  {Romanowsky} A.~J.,  {Napolitano} N.~R.,  {Arnaboldi} M.,  {Capaccioli} M.,
  {Freeman} K.~C.,  {Gerhard} O.,  {Coccato} L.,  {Carter} D.,  {Evans} N.~W.,
  {Wilkinson} M.~I.,  {Halliday} C.,    {Bridges} T.~J.,  2006, \mnras, 369,
  120

\bibitem[\protect\citeauthoryear{{Neistein}, {Maoz}, {Rix} \&
  {Tonry}}{{Neistein} et~al.}{1999}]{Neistein99}
{Neistein} E.,  {Maoz} D.,  {Rix} H.-W.,    {Tonry} J.~L.,  1999, \aj, 117,
  2666

\bibitem[\protect\citeauthoryear{{Noordermeer} \& {Verheijen}}{{Noordermeer} \&
  {Verheijen}}{2007}]{Noordermeer07c}
{Noordermeer} E.,  {Verheijen} M.~A.~W.,  2007, ArXiv Astrophysics e-prints:
  astro-ph/0708.2822

\bibitem[\protect\citeauthoryear{{Quilis}, {Moore} \& {Bower}}{{Quilis}
  et~al.}{2000}]{Quilis00}
{Quilis} V.,  {Moore} B.,    {Bower} R.,  2000, Science, 288, 1617

\bibitem[\protect\citeauthoryear{{Quinn}, {Hernquist} \& {Fullagar}}{{Quinn}
  et~al.}{1993}]{Quinn93}
{Quinn} P.~J.,  {Hernquist} L.,    {Fullagar} D.~P.,  1993, \apj, 403, 74

\bibitem[\protect\citeauthoryear{{Sancisi}, {van Woerden}, {Davies} \&
  {Hart}}{{Sancisi} et~al.}{1984}]{Sancisi84}
{Sancisi} R.,  {van Woerden} H.,  {Davies} R.~D.,    {Hart} L.,  1984, \mnras,
  210, 497

\bibitem[\protect\citeauthoryear{{Seifert} \& {Scorza}}{{Seifert} \&
  {Scorza}}{1996}]{Seifert96}
{Seifert} W.,  {Scorza} C.,  1996, \aap, 310, 75

\bibitem[\protect\citeauthoryear{{Simien} \& {Prugniel}}{{Simien} \&
  {Prugniel}}{1997}]{Simien97}
{Simien} F.,  {Prugniel} P.,  1997, \aaps, 126, 519

\bibitem[\protect\citeauthoryear{{Tonry}, {Dressler}, {Blakeslee}, {Ajhar},
  {Fletcher}, {Luppino}, {Metzger} \& {Moore}}{{Tonry} et~al.}{2001}]{Tonry01}
{Tonry} J.~L.,  {Dressler} A.,  {Blakeslee} J.~P.,  {Ajhar} E.~A.,  {Fletcher}
  A.~B.,  {Luppino} G.~A.,  {Metzger} M.~R.,    {Moore} C.~B.,  2001, \apj,
  546, 681

\bibitem[\protect\citeauthoryear{{Tully}}{{Tully}}{1980}]{Tully80}
{Tully} R.~B.,  1980, \apj, 237, 390

\bibitem[\protect\citeauthoryear{{Vel\'azquez} \& {White}}{{Vel\'azquez} \&
  {White}}{1999}]{Velazquez99}
{Vel\'azquez} H.,  {White} S.~D.~M.,  1999, \mnras, 304, 254

\end{thebibliography}

\begin{table*}
 \begin{minipage}{17.75cm}
  \centering
   \caption[Catalogue of PNe in NGC~1023 and NGC~1023A]{Catalogue of PNe in
     NGC~1023 and NGC~1023A. \label{table:catalogue}}  
  
   \begin{tabular}{lr@{\hspace{0.1cm}}r@{\hspace{0.18cm}}r@{\hspace{0.5cm}}r@{\hspace{0.1cm}}r@{\hspace{0.cm}}r@{\hspace{0.4cm}}r@{$\; \pm \;$}l@{\hspace{0.5cm}}l@{\hspace{0.75cm}}lr@{\hspace{0.1cm}}r@{\hspace{0.18cm}}r@{\hspace{0.5cm}}r@{\hspace{0.1cm}}r@{\hspace{0.cm}}r@{\hspace{0.4cm}}r@{$\; \pm \;$}l@{\hspace{0.5cm}}l}
      
    \hline
    \multicolumn{1}{c}{ID} & \multicolumn{3}{c@{\hspace{0.5cm}}}{RA (2000)} &  
    \multicolumn{3}{c@{\hspace{0.4cm}}}{Dec (2000)} & 
    \multicolumn{2}{c@{\hspace{0.6cm}}}{$V_{\mathrm {hel}}$} &
    \hspace{-0.25cm} notes & 
    \multicolumn{1}{c}{ID} & \multicolumn{3}{c@{\hspace{0.5cm}}}{RA (2000)} &
    \multicolumn{3}{c@{\hspace{0.4cm}}}{Dec (2000)} & 
    \multicolumn{2}{c@{\hspace{0.6cm}}}{$V_{\mathrm {hel}}$} &
    \hspace{-0.25cm} notes \\

    \hspace{0.2cm}PNS-EPN- & \multicolumn{1}{c@{\hspace{0.1cm}}}{\it h} &
    \multicolumn{1}{c@{\hspace{0.18cm}}}{\it m} 
    & \multicolumn{1}{c@{\hspace{0.6cm}}}{\it s} &
    \multicolumn{1}{c@{\hspace{0.1cm}}}{$^{\circ}$} & 
    \multicolumn{1}{c@{\hspace{0.cm}}}{\arcmin} & 
    \multicolumn{1}{c@{\hspace{0.3cm}}}{\arcsec} &
    \multicolumn{2}{c@{\hspace{0.5cm}}}{\hspace{-0.25cm} \kms} 
    & & \hspace{0.2cm}PNS-EPN- & \multicolumn{1}{c@{\hspace{0.1cm}}}{\it h} & 
    \multicolumn{1}{c@{\hspace{0.18cm}}}{\it m} 
    & \multicolumn{1}{c@{\hspace{0.7cm}}}{\it s} &
    \multicolumn{1}{c@{\hspace{0.1cm}}}{$^{\circ}$} & 
    \multicolumn{1}{c@{\hspace{0.cm}}}{\arcmin} & 
    \multicolumn{1}{c@{\hspace{0.3cm}}}{\arcsec} &
    \multicolumn{2}{c@{\hspace{0.5cm}}}{\hspace{-0.25cm} \kms} 
    & \\

    \hline
    NGC~1023~1  & 2 & 39 & 55.92 & 39 & 3 & 33.2  &  730 & 14 &   & NGC~1023~65  & 2 & 40 & 18.09 & 39 & 4 & 50.2  &  466 & 13 & \\   
    NGC~1023~2  & 2 & 39 & 57.40 & 39 & 3 & 14.1  &  772 & 13 &   & NGC~1023~66  & 2 & 40 & 18.18 & 39 & 3 & 55.4  &  376 & 16 & \\  
    NGC~1023~3  & 2 & 39 & 59.24 & 39 & 4 & 28.3  &  446 & 14 &   & NGC~1023~67  & 2 & 40 & 18.19 & 39 & 3 & 42.3  &  386 & 17 & \\  
    NGC~1023~4  & 2 & 40 & 1.35  & 39 & 4 & 30.6  &  639 & 13 &   & NGC~1023~68  & 2 & 40 & 18.31 & 39 & 4 & 13.9  &  348 & 14 & \\  
    NGC~1023~5  & 2 & 40 & 1.71  & 39 & 2 & 55.7  &  510 & 13 &   & NGC~1023~69  & 2 & 40 & 18.35 & 39 & 3 & 58.7  &  461 & 13 & \\  
    NGC~1023~6  & 2 & 40 & 2.55  & 39 & 4 & 17.5  &  397 & 13 &   & NGC~1023~70  & 2 & 40 & 18.42 & 39 & 3 & 28.5  &  334 & 13 & \\  
    NGC~1023~7  & 2 & 40 & 2.60  & 39 & 3 & 56.0  &  675 & 13 &   & NGC~1023~71  & 2 & 40 & 19.22 & 39 & 5 & 59.3  &  898 & 13 & \\  
    NGC~1023~8  & 2 & 40 & 3.93  & 39 & 5 & 49.4  &  473 & 14 &   & NGC~1023~72  & 2 & 40 & 19.31 & 39 & 2 & 51.7  &  494 & 14 & \\  
    NGC~1023~9  & 2 & 40 & 6.13  & 39 & 3 & 38.8  &  550 & 28 &   & NGC~1023~73  & 2 & 40 & 19.39 & 39 & 3 & 52.2  &  366 & 13 & \\  
    NGC~1023~10 & 2 & 40 & 6.48  & 39 & 2 & 56.0  &  453 & 13 &   & NGC~1023~74  & 2 & 40 & 20.20 & 39 & 3 & 16.4  &  427 & 13 & \\  
    NGC~1023~11 & 2 & 40 & 6.53  & 39 & 4 & 24.2  &  740 & 15 &   & NGC~1023~75  & 2 & 40 & 20.35 & 39 & 5 & 1.0   &  485 & 14 & \\  
    NGC~1023~12 & 2 & 40 & 6.85  & 39 & 1 & 34.9  &  767 & 13 &   & NGC~1023~76  & 2 & 40 & 20.39 & 39 & 4 & 0.9   &  561 & 18 & \\  
    NGC~1023~13 & 2 & 40 & 6.99  & 39 & 4 & 4.5   &  410 & 13 &   & NGC~1023~77  & 2 & 40 & 20.40 & 39 & 3 & 15.2  &  444 & 14 & \\  
    NGC~1023~14 & 2 & 40 & 7.00  & 39 & 1 & 17.7  &  408 & 14 &   & NGC~1023~78  & 2 & 40 & 20.97 & 39 & 3 & 49.9  &  442 & 13 & \\  
    NGC~1023~15 & 2 & 40 & 7.09  & 39 & 3 & 57.3  &  701 & 13 &   & NGC~1023~79  & 2 & 40 & 21.11 & 39 & 3 & 47.5  &  442 & 14 & \\  
    NGC~1023~16 & 2 & 40 & 7.21  & 39 & 2 & 40.2  &  446 & 13 &   & NGC~1023~80  & 2 & 40 & 21.12 & 39 & 4 & 11.7  &  543 & 13 & \\  
    NGC~1023~17 & 2 & 40 & 7.46  & 39 & 2 & 4.8   &  790 & 32 &   & NGC~1023~81  & 2 & 40 & 21.22 & 39 & 3 & 8.3   &  535 & 16 & \\  
    NGC~1023~18 & 2 & 40 & 8.38  & 39 & 3 & 33.8  &  456 & 16 &   & NGC~1023~82  & 2 & 40 & 21.81 & 39 & 1 & 11.6  &  787 & 18 & \\  
    NGC~1023~19 & 2 & 40 & 8.46  & 39 & 3 & 43.0  &  548 & 13 &   & NGC~1023~83  & 2 & 40 & 21.85 & 39 & 5 & 9.1   &  619 & 13 & \\  
    NGC~1023~20 & 2 & 40 & 9.68  & 39 & 2 & 50.2  &  379 & 13 &   & NGC~1023~84  & 2 & 40 & 21.95 & 39 & 5 & 25.9  &  603 & 13 & \\  
    NGC~1023~21 & 2 & 40 & 9.77  & 39 & 3 & 10.2  &  385 & 13 &   & NGC~1023~85  & 2 & 40 & 22.26 & 39 & 2 & 31.7  &  606 & 13 & \\  
    NGC~1023~22 & 2 & 40 & 9.81  & 39 & 3 & 13.9  &  447 & 13 &   & NGC~1023~86  & 2 & 40 & 22.27 & 39 & 2 & 3.8   &  661 & 15 & \\  
    NGC~1023~23 & 2 & 40 & 10.23 & 39 & 3 & 36.8  &  733 & 13 & a & NGC~1023~87  & 2 & 40 & 22.28 & 39 & 2 & 51.2  &  537 & 13 & \\  
    NGC~1023~24 & 2 & 40 & 10.91 & 39 & 4 & 29.6  &  372 & 13 &   & NGC~1023~88  & 2 & 40 & 22.31 & 39 & 4 & 20.2  &  592 & 13 & \\  
    NGC~1023~25 & 2 & 40 & 11.07 & 39 & 3 & 53.3  &  489 & 13 &   & NGC~1023~89  & 2 & 40 & 22.32 & 39 & 4 & 29.5  &  586 & 14 & \\  
    NGC~1023~26 & 2 & 40 & 11.11 & 39 & 2 & 25.5  &  600 & 14 &   & NGC~1023~90  & 2 & 40 & 22.60 & 39 & 4 & 39.9  &  696 & 13 & \\  
    NGC~1023~27 & 2 & 40 & 11.40 & 39 & 2 & 23.8  &  587 & 23 &   & NGC~1023~91  & 2 & 40 & 22.64 & 39 & 4 & 13.9  &  485 & 14 & \\  
    NGC~1023~28 & 2 & 40 & 11.64 & 39 & 4 & 50.6  &  511 & 13 &   & NGC~1023~92  & 2 & 40 & 23.06 & 39 & 4 & 9.2   &  675 & 14 & \\  
    NGC~1023~29 & 2 & 40 & 11.84 & 39 & 3 & 49.4  &  465 & 13 &   & NGC~1023~93  & 2 & 40 & 23.51 & 39 & 3 & 26.8  &  599 & 15 & \\  
    NGC~1023~30 & 2 & 40 & 12.16 & 39 & 3 & 32.8  &  490 & 13 &   & NGC~1023~94  & 2 & 40 & 23.84 & 39 & 4 & 31.8  &  499 & 13 & \\  
    NGC~1023~31 & 2 & 40 & 12.32 & 39 & 3 & 44.0  &  350 & 19 &   & NGC~1023~95  & 2 & 40 & 24.05 & 39 & 2 & 47.4  &  582 & 31 & \\  
    NGC~1023~32 & 2 & 40 & 12.78 & 39 & 3 & 6.6   &  511 & 24 &   & NGC~1023~96  & 2 & 40 & 24.22 & 39 & 5 & 27.5  &  768 & 13 & \\  
    NGC~1023~33 & 2 & 40 & 12.89 & 39 & 3 & 7.5   &  523 & 24 &   & NGC~1023~97  & 2 & 40 & 24.41 & 39 & 3 & 15.2  &  742 & 14 & \\  
    NGC~1023~34 & 2 & 40 & 13.49 & 39 & 4 & 30.2  &  471 & 13 &   & NGC~1023~98  & 2 & 40 & 24.49 & 39 & 8 & 11.0  &  737 & 14 & \\  
    NGC~1023~35 & 2 & 40 & 13.52 & 39 & 3 & 58.6  &  447 & 14 &   & NGC~1023~99  & 2 & 40 & 25.05 & 39 & 4 & 47.8  &  738 & 14 & \\  
    NGC~1023~36 & 2 & 40 & 13.79 & 39 & 3 & 28.5  &  390 & 14 &   & NGC~1023~100 & 2 & 40 & 25.06 & 39 & 4 & 22.4  &  588 & 14 & \\  
    NGC~1023~37 & 2 & 40 & 14.12 & 39 & 5 & 45.2  &  566 & 13 &   & NGC~1023~101 & 2 & 40 & 25.24 & 39 & 4 & 7.8   &  512 & 17 & \\  
    NGC~1023~38 & 2 & 40 & 14.25 & 39 & 3 & 16.3  &  305 & 16 &   & NGC~1023~102 & 2 & 40 & 25.31 & 39 & 3 & 16.9  &  612 & 14 & \\  
    NGC~1023~39 & 2 & 40 & 14.57 & 39 & 4 & 36.0  &  479 & 13 &   & NGC~1023~103 & 2 & 40 & 25.33 & 39 & 4 & 20.9  &  571 & 13 & \\  
    NGC~1023~40 & 2 & 40 & 14.61 & 39 & 3 & 18.2  &  357 & 13 &   & NGC~1023~104 & 2 & 40 & 25.48 & 39 & 4 & 22.7  &  729 & 14 & \\  
    NGC~1023~41 & 2 & 40 & 14.65 & 39 & 3 & 54.5  &  371 & 13 &   & NGC~1023~105 & 2 & 40 & 25.59 & 39 & 4 & 53.2  &  745 & 13 & \\  
    NGC~1023~42 & 2 & 40 & 14.85 & 39 & 3 & 3.5   &  441 & 13 &   & NGC~1023~106 & 2 & 40 & 25.78 & 39 & 3 & 1.5   &  531 & 14 & \\  
    NGC~1023~43 & 2 & 40 & 14.90 & 39 & 4 & 16.0  &  898 & 13 & a & NGC~1023~107 & 2 & 40 & 25.80 & 39 & 3 & 32.0  &  534 & 13 & \\  
    NGC~1023~44 & 2 & 40 & 14.97 & 39 & 3 & 16.5  &  449 & 13 &   & NGC~1023~108 & 2 & 40 & 25.83 & 39 & 3 & 35.1  &  573 & 16 & \\  
    NGC~1023~45 & 2 & 40 & 15.11 & 39 & 3 & 45.0  &  516 & 13 &   & NGC~1023~109 & 2 & 40 & 25.94 & 39 & 3 & 9.6   &  591 & 13 & \\  
    NGC~1023~46 & 2 & 40 & 15.11 & 39 & 4 & 30.8  &  560 & 13 &   & NGC~1023~110 & 2 & 40 & 26.03 & 39 & 2 & 42.6  &  688 & 13 & \\  
    NGC~1023~47 & 2 & 40 & 15.25 & 39 & 3 & 9.1   &  372 & 13 &   & NGC~1023~111 & 2 & 40 & 26.05 & 39 & 4 & 30.6  &  672 & 13 & \\  
    NGC~1023~48 & 2 & 40 & 15.33 & 39 & 2 & 55.0  &  346 & 13 &   & NGC~1023~112 & 2 & 40 & 26.22 & 39 & 2 & 19.8  &  566 & 13 & \\  
    NGC~1023~49 & 2 & 40 & 16.03 & 39 & 3 & 0.6   &  486 & 13 &   & NGC~1023~113 & 2 & 40 & 26.26 & 39 & 2 & 33.6  &  662 & 13 & \\  
    NGC~1023~50 & 2 & 40 & 16.17 & 39 & 3 & 7.8   &  545 & 13 &   & NGC~1023~114 & 2 & 40 & 26.39 & 39 & 4 & 47.6  &  681 & 13 & \\  
    NGC~1023~51 & 2 & 40 & 16.31 & 39 & 4 & 22.8  &  440 & 13 &   & NGC~1023~115 & 2 & 40 & 26.82 & 39 & 3 & 17.4  &  691 & 14 & \\  
    NGC~1023~52 & 2 & 40 & 16.33 & 39 & 2 & 54.8  &  581 & 13 &   & NGC~1023~116 & 2 & 40 & 26.92 & 39 & 4 & 2.8   &  928 & 13 & \\  
    NGC~1023~53 & 2 & 40 & 16.35 & 39 & 2 & 38.8  &  775 & 13 &   & NGC~1023~117 & 2 & 40 & 27.01 & 39 & 3 & 46.0  &  740 & 13 & \\  
    NGC~1023~54 & 2 & 40 & 16.41 & 39 & 2 & 6.0   &  381 & 16 &   & NGC~1023~118 & 2 & 40 & 27.08 & 39 & 4 & 12.6  &  767 & 13 & \\  
    NGC~1023~55 & 2 & 40 & 16.49 & 39 & 4 & 57.3  &  474 & 14 &   & NGC~1023~119 & 2 & 40 & 27.12 & 39 & 3 & 51.0  &  854 & 17 & \\  
    NGC~1023~56 & 2 & 40 & 16.60 & 39 & 3 & 19.1  &  460 & 13 &   & NGC~1023~120 & 2 & 40 & 27.33 & 39 & 4 & 1.5   &  724 & 13 & \\  
    NGC~1023~57 & 2 & 40 & 16.86 & 39 & 2 & 19.4  &  570 & 19 &   & NGC~1023~121 & 2 & 40 & 27.35 & 39 & 4 & 8.4   &  801 & 13 & \\  
    NGC~1023~58 & 2 & 40 & 17.05 & 39 & 1 & 2.3   &  591 & 13 &   & NGC~1023~122 & 2 & 40 & 27.48 & 39 & 4 & 35.5  &  736 & 13 & \\  
    NGC~1023~59 & 2 & 40 & 17.17 & 39 & 3 & 0.3   &  467 & 14 &   & NGC~1023~123 & 2 & 40 & 27.51 & 39 & 4 & 39.0  &  794 & 13 & \\  
    NGC~1023~60 & 2 & 40 & 17.30 & 39 & 4 & 3.0   &  389 & 14 &   & NGC~1023~124 & 2 & 40 & 27.52 & 39 & 3 & 10.9  &  625 & 18 & \\  
    NGC~1023~61 & 2 & 40 & 17.36 & 39 & 4 & 11.0  &  540 & 13 &   & NGC~1023~125 & 2 & 40 & 27.60 & 39 & 4 & 15.5  &  846 & 13 & \\  
    NGC~1023~62 & 2 & 40 & 17.56 & 39 & 3 & 47.3  &  635 & 13 &   & NGC~1023~126 & 2 & 40 & 28.11 & 39 & 4 & 6.0   &  802 & 15 & \\  
    NGC~1023~63 & 2 & 40 & 17.56 & 39 & 4 & 9.4   &  338 & 14 &   & NGC~1023~127 & 2 & 40 & 28.18 & 39 & 2 & 3.1   &  410 & 14 & \\   
    NGC~1023~64 & 2 & 40 & 17.95 & 39 & 4 & 10.0  &  495 & 13 &   & NGC~1023~128 & 2 & 40 & 28.65 & 39 & 3 & 54.8  &  773 & 13 & \\ 
    \hline
   \end{tabular}
 \end{minipage}
\end{table*}  
  
\begin{table*}
 \begin{minipage}{17.75cm}
  \centering
   \contcaption{Catalogue of PNe in NGC~1023 and NGC~1023A.}  
  
   \begin{tabular}{lr@{\hspace{0.1cm}}r@{\hspace{0.18cm}}r@{\hspace{0.5cm}}r@{\hspace{0.1cm}}r@{\hspace{0.cm}}r@{\hspace{0.4cm}}r@{$\; \pm \;$}l@{\hspace{0.5cm}}l@{\hspace{0.75cm}}lr@{\hspace{0.1cm}}r@{\hspace{0.18cm}}r@{\hspace{0.5cm}}r@{\hspace{0.1cm}}r@{\hspace{0.cm}}r@{\hspace{0.4cm}}r@{$\; \pm \;$}l@{\hspace{0.5cm}}l}
      
    \hline
    \multicolumn{1}{c}{ID} & \multicolumn{3}{c@{\hspace{0.5cm}}}{RA (2000)} &  
    \multicolumn{3}{c@{\hspace{0.4cm}}}{Dec (2000)} & 
    \multicolumn{2}{c@{\hspace{0.6cm}}}{$V_{\mathrm {hel}}$} &
    \hspace{-0.25cm} notes & 
    \multicolumn{1}{c}{ID} & \multicolumn{3}{c@{\hspace{0.5cm}}}{RA (2000)} &
    \multicolumn{3}{c@{\hspace{0.4cm}}}{Dec (2000)} & 
    \multicolumn{2}{c@{\hspace{0.6cm}}}{$V_{\mathrm {hel}}$} &
    \hspace{-0.25cm} notes \\

    \hspace{0.2cm}PNS-EPN- & \multicolumn{1}{c@{\hspace{0.1cm}}}{\it h} &
    \multicolumn{1}{c@{\hspace{0.18cm}}}{\it m} 
    & \multicolumn{1}{c@{\hspace{0.6cm}}}{\it s} &
    \multicolumn{1}{c@{\hspace{0.1cm}}}{$^{\circ}$} & 
    \multicolumn{1}{c@{\hspace{0.cm}}}{\arcmin} & 
    \multicolumn{1}{c@{\hspace{0.3cm}}}{\arcsec} &
    \multicolumn{2}{c@{\hspace{0.5cm}}}{\hspace{-0.25cm} \kms} 
    & & \hspace{0.2cm}PNS-EPN- & \multicolumn{1}{c@{\hspace{0.1cm}}}{\it h} & 
    \multicolumn{1}{c@{\hspace{0.18cm}}}{\it m} 
    & \multicolumn{1}{c@{\hspace{0.7cm}}}{\it s} &
    \multicolumn{1}{c@{\hspace{0.1cm}}}{$^{\circ}$} & 
    \multicolumn{1}{c@{\hspace{0.cm}}}{\arcmin} & 
    \multicolumn{1}{c@{\hspace{0.3cm}}}{\arcsec} &
    \multicolumn{2}{c@{\hspace{0.5cm}}}{\hspace{-0.25cm} \kms} 
    & \\

    \hline      
    NGC~1023~129 & 2 & 40 & 28.77 & 39 & 3 & 47.1  &  684 & 13 &   & NGC~1023~167 & 2 & 40 & 35.70 & 39 & 3 & 45.2  &  753 & 13 &   \\ 
    NGC~1023~130 & 2 & 40 & 28.87 & 39 & 4 & 0.2   &  846 & 13 &   & NGC~1023~168 & 2 & 40 & 35.74 & 39 & 3 & 26.3  &  830 & 13 & c \\ 
    NGC~1023~131 & 2 & 40 & 29.07 & 39 & 3 & 22.5  &  859 & 13 &   & NGC~1023~169 & 2 & 40 & 35.76 & 39 & 3 & 32.8  &  777 & 14 & c \\ 
    NGC~1023~132 & 2 & 40 & 29.20 & 39 & 4 & 35.7  &  762 & 13 &   & NGC~1023~170 & 2 & 40 & 36.76 & 39 & 3 & 14.6  &  723 & 13 & c \\ 
    NGC~1023~133 & 2 & 40 & 29.31 & 39 & 3 & 42.2  &  726 & 13 &   & NGC~1023~171 & 2 & 40 & 36.83 & 39 & 3 & 46.9  &  767 & 13 & c \\ 
    NGC~1023~134 & 2 & 40 & 29.38 & 39 & 4 & 17.0  &  725 & 16 &   & NGC~1023~172 & 2 & 40 & 36.86 & 39 & 2 & 52.3  &  741 & 13 & c \\ 
    NGC~1023~135 & 2 & 40 & 29.39 & 39 & 4 & 21.9  &  859 & 14 &   & NGC~1023~173 & 2 & 40 & 37.07 & 39 & 3 & 3.6   &  771 & 14 & c \\ 
    NGC~1023~136 & 2 & 40 & 29.43 & 39 & 3 & 28.4  &  745 & 14 &   & NGC~1023~174 & 2 & 40 & 37.44 & 39 & 3 & 23.7  &  732 & 13 & c \\ 
    NGC~1023~137 & 2 & 40 & 29.49 & 39 & 3 & 5.4   &  765 & 13 &   & NGC~1023~175 & 2 & 40 & 37.55 & 39 & 3 & 43.2  &  722 & 13 & c \\ 
    NGC~1023~138 & 2 & 40 & 29.53 & 39 & 4 & 45.7  &  683 & 14 &   & NGC~1023~176 & 2 & 40 & 37.79 & 39 & 3 & 41.8  &  828 & 13 & c \\ 
    NGC~1023~139 & 2 & 40 & 29.65 & 39 & 3 & 58.8  &  649 & 13 &   & NGC~1023~177 & 2 & 40 & 37.87 & 39 & 2 & 25.6  &  750 & 13 & c \\ 
    NGC~1023~140 & 2 & 40 & 29.72 & 39 & 2 & 23.8  &  741 & 13 &   & NGC~1023~178 & 2 & 40 & 38.01 & 39 & 3 & 1.5   &  744 & 13 & c \\ 
    NGC~1023~141 & 2 & 40 & 30.06 & 39 & 4 & 0.0   &  833 & 13 &   & NGC~1023~179 & 2 & 40 & 38.13 & 39 & 3 & 35.5  &  808 & 13 & c \\  
    NGC~1023~142 & 2 & 40 & 30.25 & 39 & 3 & 51.4  &  835 & 13 &   & NGC~1023~180 & 2 & 40 & 38.16 & 39 & 4 & 14.2  &  748 & 13 &   \\  
    NGC~1023~143 & 2 & 40 & 30.44 & 39 & 4 & 3.8   &  661 & 17 &   & NGC~1023~181 & 2 & 40 & 38.79 & 39 & 3 & 34.8  &  848 & 13 & c \\  
    NGC~1023~144 & 2 & 40 & 30.58 & 39 & 5 & 54.7  &  450 & 13 &   & NGC~1023~182 & 2 & 40 & 39.41 & 39 & 4 & 49.7  &  750 & 13 &   \\  
    NGC~1023~145 & 2 & 40 & 30.60 & 39 & 3 & 39.0  &  855 & 13 &   & NGC~1023~183 & 2 & 40 & 39.91 & 39 & 4 & 12.1  &  822 & 13 & c \\  
    NGC~1023~146 & 2 & 40 & 30.74 & 39 & 3 & 20.3  &  812 & 13 &   & NGC~1023~184 & 2 & 40 & 40.04 & 39 & 3 & 45.9  &  736 & 13 & c \\  
    NGC~1023~147 & 2 & 40 & 31.08 & 39 & 4 & 38.2  &  868 & 13 &   & NGC~1023~185 & 2 & 40 & 40.24 & 39 & 5 & 29.0  &  716 & 13 &   \\  
    NGC~1023~148 & 2 & 40 & 31.29 & 39 & 3 & 27.2  &  809 & 13 &   & NGC~1023~186 & 2 & 40 & 40.52 & 39 & 7 & 1.3   &  664 & 13 &   \\  
    NGC~1023~149 & 2 & 40 & 31.76 & 39 & 3 & 34.2  &  766 & 14 &   & NGC~1023~187 & 2 & 40 & 40.71 & 39 & 3 & 50.4  &  781 & 13 & c \\  
    NGC~1023~150 & 2 & 40 & 31.77 & 39 & 3 & 41.1  &  874 & 13 &   & NGC~1023~188 & 2 & 40 & 40.74 & 39 & 5 & 51.1  &  769 & 13 &   \\  
    NGC~1023~151 & 2 & 40 & 32.48 & 39 & 3 & 54.6  &  772 & 14 &   & NGC~1023~189 & 2 & 40 & 40.78 & 39 & 4 & 38.4  &  787 & 13 &   \\  
    NGC~1023~152 & 2 & 40 & 32.77 & 39 & 4 & 11.8  &  856 & 13 &   & NGC~1023~190 & 2 & 40 & 40.85 & 39 & 6 & 33.6  &  700 & 35 &   \\  
    NGC~1023~153 & 2 & 40 & 32.85 & 39 & 4 & 49.2  &  745 & 13 &   & NGC~1023~191 & 2 & 40 & 41.09 & 39 & 4 & 21.1  &  419 & 13 & a \\  
    NGC~1023~154 & 2 & 40 & 33.21 & 39 & 3 & 37.7  &  776 & 13 &   & NGC~1023~192 & 2 & 40 & 41.41 & 39 & 5 & 17.1  &  717 & 14 &   \\ 
    NGC~1023~155 & 2 & 40 & 33.30 & 39 & 3 & 29.8  &  780 & 13 &   & NGC~1023~193 & 2 & 40 & 41.67 & 39 & 8 & 17.0  &  610 & 13 &   \\  
    NGC~1023~156 & 2 & 40 & 33.35 & 39 & 3 & 47.0  &  823 & 16 &   & NGC~1023~194 & 2 & 40 & 41.89 & 39 & 1 & 28.3  &  757 & 13 &   \\  
    NGC~1023~157 & 2 & 40 & 33.39 & 39 & 3 & 10.6  &  786 & 14 &   & NGC~1023~195 & 2 & 40 & 42.40 & 39 & 3 & 43.5  &  748 & 13 & c \\  
    NGC~1023~158 & 2 & 40 & 33.68 & 39 & 4 & 46.5  &  728 & 13 &   & NGC~1023~196 & 2 & 40 & 42.54 & 39 & 4 & 25.3  &  813 & 16 &   \\  
    NGC~1023~159 & 2 & 40 & 33.68 & 39 & 3 & 24.7  &  765 & 15 &   & NGC~1023~197 & 2 & 40 & 43.07 & 39 & 2 & 38.0  &  741 & 13 &   \\  
    NGC~1023~160 & 2 & 40 & 33.76 & 39 & 3 & 10.5  &  810 & 14 & c & NGC~1023~198 & 2 & 40 & 43.09 & 39 & 4 & 4.8   &  658 & 13 &   \\  
    NGC~1023~161 & 2 & 40 & 34.04 & 39 & 4 & 35.2  &  618 & 13 &   & NGC~1023~199 & 2 & 40 & 44.40 & 39 & 3 & 43.5  &  785 & 13 &   \\  
    NGC~1023~162 & 2 & 40 & 34.40 & 39 & 2 & 37.0  &  755 & 14 & c & NGC~1023~200 & 2 & 40 & 44.73 & 39 & 2 & 7.0   & 1703 & 13 & b \\  
    NGC~1023~163 & 2 & 40 & 34.50 & 39 & 4 & 2.9   &  813 & 13 &   & NGC~1023~201 & 2 & 40 & 44.77 & 39 & 4 & 7.6   &  545 & 13 &   \\  
    NGC~1023~164 & 2 & 40 & 34.68 & 39 & 3 & 57.4  &  832 & 17 &   & NGC~1023~202 & 2 & 40 & 46.40 & 39 & 1 & 3.4   &  452 & 13 &   \\  
    NGC~1023~165 & 2 & 40 & 34.83 & 39 & 3 & 43.4  &  812 & 13 &   & NGC~1023~203 & 2 & 40 & 47.71 & 39 & 6 & 5.1   &  804 & 13 &   \\  
    NGC~1023~166 & 2 & 40 & 35.53 & 39 & 3 & 12.3  &  745 & 13 & c & NGC~1023~204 & 2 & 40 & 48.98 & 39 & 3 & 58.8  &  734 & 13 &   \\  
    \hline
    \multicolumn{20}{l}{explanation of the notes:} \\
    \multicolumn{20}{l}{a \hspace{0.15cm} $> 3 \sigma$ outlier in the tilted ring analysis} \\
    \multicolumn{20}{l}{b \hspace{0.15cm} Discrepant velocity; unbound to system} \\
    \multicolumn{20}{l}{c \hspace{0.15cm} Object statistically more likely to
    belong to companion NGC~1023A than to main galaxy NGC~1023.} \\  
   \end{tabular}
 \end{minipage}
\end{table*}

\end{document}